\documentstyle[preprint,aps]{revtex}
\begin{document}
\input epsf
\title{Kondo effect in Normal-Superconductor Quantum Dots}

\author{J.C. Cuevas$^{1,2}$, A. Levy Yeyati$^1$, A. Mart\'{\i}n-Rodero$^1$}

\address{$^1$ Departamento de F\'\i sica Te\'orica de la Materia Condensada 
C-V, Universidad Aut\'onoma de Madrid, E-28049 Madrid, Spain \\
$^2$ Institut f\"ur Theoretische Festk\"orperphysik, Universit\"at
Karlsruhe, 76128 Karlsruhe, Germany}

\date{\today}

\maketitle

\begin{abstract}
We study the transport properties of a quantum dot coupled to a normal
and a superconducting lead. The dot is represented by a generalized
Anderson model. 
Correlation effects are taken into account by
an appropriate self-energy which interpolates between the
limits of weak and strong coupling to the leads. 
The transport properties of the system are controlled by the 
interplay between
the Kondo effect and Andreev reflection processes. We show that,
depending on the parameters range the conductance can either be enhanced
or suppressed as compared to the normal case. In particular, by adequately
tunning the coupling to the leads one can reach the maximum value
$4e^2/h$ for the conductance. 
\end{abstract}



\narrowtext

The Kondo effect is a prototypical correlation effect in Solid State
Physics. Although it was first analyzed for the case of 
magnetic impurities in metals,
in the last years there has been a renewed interest in Kondo physics
with its observation in a semiconductor quantum dot (QD)
\cite{Goldhaber,Kowenhoven}. Quantum dots
constitute an ideal laboratory for testing the theoretical predictions
as they allow to vary the relevant parameters in the problem
in a controlled way. This technology also opens some new 
possibilities like the exploration of the Kondo effect when 
the dot is connected to superconducting leads. The interesting issue 
in this case is related to
the competition between the strong Coulomb interaction 
in the quantum dot and the pairing interaction within the leads.

From the theoretical side, the Kondo effect in QDs has been mainly analyzed
by means of the single-level Anderson model \cite{Glazman,us}. 
The theory predicts an enhancement of the dot conductance at low
temperatures due to the development of the so-called Kondo resonance. 
The case when
one of the leads is superconducting has been recently analyzed by some
authors using different theoretical methods \cite{Fazio,Kang,Ambegaokar}
assuming a modified Anderson model in which one of the metallic
electrodes is substituted by a BCS superconductor. 
While some authors have predicted an enhancement of the conductance 
due to Andreev reflection at
the superconducting lead \cite{Kang}, others have predicted 
the opposite effect \cite{Ambegaokar}. 
In Refs. \cite{Fazio,Ambegaokar} the infinite charging energy limit ($U
\rightarrow \infty$) has been assumed. However, in an actual experiment
this assumption may not be completely justified (for instance, in the
experiments of Ref. \cite{Goldhaber} the ratio $U/\Gamma$, $\Gamma$
being the dot tunneling rate, was estimated to be around 6.5). 
The approach presented in this letter would allow
to analyze this problem for a broad range of the different parameters  
of the model. We will show that, depending on the values of these
parameters, one can obtain either an enhancement or a reduction of the
conductance with respect to the normal case.

Our approximation scheme is based on the hypothesis that a good
approximation to the electron self-energy can be found by interpolating 
between the limits of weak and strong coupling to the leads. This
interpolative method has been applied successfully to analyze different
strongly correlated electron systems like the equilibrium \cite{Alvaro,Saso},
the non-equilibrium \cite{us} and the multilevel \cite{us2} Anderson models
and the Hubbard model \cite{Alvaro2,Kotliar}.
In this letter we shall discuss how to extend this method to the
superconducting case. 

For describing a N-QD-S system we use an Anderson-like Hamiltonian

\begin{equation}
\hat{H} = \hat{H}_N + \hat{H}_S + \sum_{\sigma} \epsilon_0 \hat{n}_{\sigma}
+ U \hat{n}_{\uparrow} \hat{n}_{\downarrow} + \hat{H}_T,
\end{equation}
where $\hat{n}_{\sigma}=\hat{d}^{\dagger}_{\sigma} \hat{d}_{\sigma}$,
$\hat{H}_N$ and $\hat{H}_S$ represent the uncoupled normal and
superconducting leads respectively; $\hat{H}_T = \sum_{k \in N,S;\sigma}
t_{0,k} \hat{d}^{\dagger}_{\sigma} \hat{c}_{k,\sigma} + h.c.$
describing the coupling between the dot level and the leads. Within
this model the dot is represented by a single spin degenerate level
with a repulsive Coulomb interaction described by the U-term in Eq. (1).
We shall assume that the superconducting lead is well described by the
BCS theory with a superconducting gap $\Delta$ and the normal lead is,
as usual, characterized by a flat density of states around the Fermi 
level, $\rho_F$.

The transport properties of this model can be obtained by means of Green
function techniques. In order to analyze the linear regime the main 
quantity to be determined is the dot retarded Green function, which in 
a Nambu $2 \times 2$ representation adopts the form

\begin{equation}
\hat{G}^r(\omega) = \left[ \omega \hat{I} - \epsilon_0 \hat{\sigma}_z -
\hat{\Sigma}^r(\omega) - \hat{\Gamma}_N (\omega) - \hat{\Gamma}_S
(\omega) \right]^{-1},
\end{equation} 
where $\hat{\Gamma}_N$ and $\hat{\Gamma}_S$ are the tunneling rates
given by $\hat{\Gamma}_N = \Gamma_L \hat{I}$,
and the superconducting tunneling rate
$\hat{\Gamma}_S$ is given by $\hat{\Gamma}_S = \Gamma_R \hat{g}$,
where $\Gamma_{L,R} = \pi t^2_{L,R} \rho_F$,
$g_{11} = g_{22} = - \omega/\sqrt{\Delta^2 - \omega^2}$ and
$g_{12} = g_{21} = \Delta/\sqrt{\Delta^2 - \omega^2}$ (the chemical
potential of the superconducting lead is taken as zero). The self-energy
$\hat{\Sigma}^r(\omega)$ takes into account the effect of Coulomb 
interactions.
To the lowest order in $U$ this is given by the Hartree-Fock Bogoliubov
approximation: $\hat{\Sigma}^r = U <\hat{n}> \hat{\sigma}_z + \Delta_d
\hat{\sigma}_x$, $\Delta_d$ being the proximity effect
induced order parameter in the QD,
$\Delta_d = U \langle \hat{d}^{\dagger}_{\uparrow}
\hat{d}^{\dagger}_{\downarrow} \rangle$. 
The crucial problem is to find a good approximation to include correlation 
effects beyond this mean field approximation. 

Within the spirit of the interpolative method commented above,
the self-energy is constructed in such a way as to interpolate 
between the limits of weak and strong coupling to the leads, for which the
exact result are known. Let us first analyze the weak coupling or {\it
atomic} limit. In this case we have $t_{L,R}/U \rightarrow 0$
and thus $\hat{\Gamma}_N/U$, $\hat{\Gamma}_S/U$ $\rightarrow 0$. In this
limit the induced order parameter in the QD vanishes faster than $(t_R/U)^2$
and one can neglect the non-diagonal elements in the self-energy matrix.
On the other hand, the diagonal elements can be easily evaluated in this
limit using the equation of motion method \cite{Alvaro} and have the form
\begin{equation}
\Sigma^r_{11,22} \rightarrow \pm U <\hat{n}> + 
\frac{U^2 <\hat{n}>(1 - <\hat{n}>)}{\omega \mp \epsilon_0
\mp U (1 - <\hat{n}>)} 
\end{equation}

In the opposite limit, $U/t_{L,R} \rightarrow 0$, one can accurately
evaluate the self-energy using standard perturbation theory in the
Coulomb interaction. The different diagrams
contributing to the second order self-energy are depicted
in Fig 1. In the superconducting case, there appear
additional diagrams to the one in the normal case
(diagram a) corresponding to the interaction of an electron with an
electron-hole pair in the QD; the remaining diagrams contain at least
one anomalous propagator and vanish identically in the normal state.
As in the normal case \cite{us}, the non-perturbed one-electron Hamiltonian, 
over which the diagrammatic series is constructed, is taken as an
effective mean field, characterized by an effective dot level
$\epsilon_{eff}$, having the same dot charge as the fully interacting
problem. As shown in Ref. \cite{us} this self-consistency condition
provides in the normal case a good fulfillment of the Friedel sum rule 
at zero temperature. 
The extension of this procedure to the superconducting case requires
dressing the propagators in the diagrams of Fig. 1 with the non-diagonal
self-energy $\Sigma_{12}$ in order to impose 
also consistency in the non-diagonal 
charge $\langle \hat{d}^{\dagger}_{\uparrow} \hat{d}^{\dagger}_{\downarrow} 
\rangle$. 
Notice that although the interaction in the QD is repulsive, there is
always some induced paring potential in the dot due to the proximity
effect. The inclusion of this effect for finite $U$ 
is very important for the correct description of the dot electronic 
properties. 

The original interpolative scheme stems from the observation 
that the second order self-energy ($\Sigma^{(2)}$) has a similar
functional form as the atomic self-energy for large frequencies 
\cite{Alvaro} thus allowing for a smooth interpolation between
the two limits. 
In the superconducting case, the diagonal
elements of the second order self-energy behave as 
\begin{equation}
\Sigma^{(2)}_{11,22} \sim \frac{U^2 <\hat{n}>(1 - <\hat{n}>)}{\omega \mp
\epsilon_{eff}} 
\end{equation}
for large frequencies, while the non-diagonal elements decay faster than
$U^2/\omega$. This behavior permits to define a Nambu $2 \times 2$
interpolative ansatz for the self-energy matrix as:
\begin{equation}
\hat{\Sigma}(\omega) = U \langle \hat{n} \rangle \hat{\sigma}_z + \Delta_d 
\hat{\sigma}_x + \left[\hat{I} - \alpha \hat{\Sigma}^{(2)}
\hat{\sigma}_z \right]^{-1} \hat{\Sigma}^{(2)}(\omega) ,
\end{equation}
where 
\[\alpha = \frac{\epsilon_0 + (1-<\hat{n}>)U -\epsilon_{eff}}{U^2
<\hat{n}>(1-<\hat{n}>)}. \]

\noindent
and $\hat{\Sigma}^{(2)}$ is the second order self-energy matrix whose
elements are given by the diagrams depicted in Fig. 1.

Using this ansatz one recovers the correct behavior of the self-energy
both in the weak and strong coupling limits. Moreover, this ansatz
satisfies the exact relations between the different matrix elements, i.e
$\Sigma_{12}(\omega) = \Sigma_{21}(\omega)$ and $\Sigma_{11}(\omega) =
-\Sigma^{*}_{22}(- \omega)$.

Due to the presence of an additional energy scale fixed by $\Delta$ 
the number of different physical regimes is larger than in the normal case. 
We will mainly consider the more interesting physical regime $\Gamma =
\Gamma_L + \Gamma_R \sim \Delta$ \cite{comment}.
In Fig. 2a we show the dot spectral density (LDOS) 
for a symmetric case ($\epsilon_0 = -U/2$) with 
$\Gamma_L=\Gamma_R=\Delta$ and increasing values of $U$.
As can be observed, when $U \le \Delta$ the LDOS exhibits a double peak
around the Fermi energy which is due to the influence of the
superconducting electrode by the proximity effect. However, as $U$
increases the double peak is replaced by a single narrow Kondo resonance 
as in the normal case. The comparison with the normal case
reveals that the Kondo resonance gets narrower in
the superconducting case and its height increases with $U$ above the 
normal value. In the limit $U \rightarrow \infty$, this height
approaches the value $2/(\pi \Gamma)$, which is twice the value 
in the normal case at zero temperature, as fixed by the 
Friedel sum rule. The narrowing of the Kondo resonance gives rise to a 
lowering of the Kondo temperature with respect to the normal case.

For energies larger than 
$\Delta$ the differences between the normal and the superconducting LDOS
become negligible, with the usual broad resonances at $\epsilon_0$ and
$\epsilon_0 + U$ which become more pronounced for increasing $U$.  

By varying the dot level position $\epsilon_0$ one can study the
transition from the Kondo to the mixed valence regime. The evolution of
the dot LDOS is illustrated in Fig. 2b. When approaching the mixed
valence regime ($|\epsilon_0| < \Gamma$ or $|\epsilon_0 + U| < \Gamma$) 
the Kondo resonance is replaced
by an asymmetric broad resonance close to the Fermi energy as in the normal
case. In the superconducting case, however, the LDOS develops an
additional structure associated with the BCS divergencies at the gap
edges \cite{Ambegaokar}. 

As in any NS contact, transport at low voltages is possible 
due to Andreev reflection processes. At finite temperature, the linear
conductance is given by the expression \cite{Raimondi}

\begin{equation}
G = \frac{16e^2}{h} \Gamma_L \int^{\infty}_{-\infty} dE \; 
\mbox{Im} \left( G^r_{12} G^a_{11} \right) \left( \Gamma_R - \mbox{Re}
\Sigma_{12} \right) \left( - \frac{ \partial f}{\partial E} \right),
\label{conductance}
\end{equation}
where $f(E)$ is the Fermi function.
At zero temperature, $\mbox{Im}\hat{\Sigma}(0)=0$, and
Eq. (\ref{conductance}) reduces to

\begin{equation}
G = \frac{4e^2}{h} \frac{4 \Gamma_L^2
\tilde{\Gamma}_R^2}{\left[\tilde{\epsilon}^2 + \Gamma_L^2 + \tilde{\Gamma}_R^2
\right]^2}, 
\label{cond0}
\end{equation}
where $\tilde{\Gamma}_R = \Gamma_R - \mbox{Re}\Sigma_{12}(0)$ and 
$\tilde{\epsilon} = \epsilon_0 + \mbox{Re} \Sigma_{11}(0)$. Notice that
Eq. (\ref{cond0}) coincides at $U=0$ with the well known non-interacting
result \cite{Beenakker}.

One would expect that for a dot symmetrically coupled to the leads (i.e.
$\Gamma_L = \Gamma_R$) and in the case of electron-hole symmetry
($\epsilon_0 = -U/2$),
the conductance should reach its maximum value $4 e^2/h$ \cite{Kang}. 
However, the actual situation is more complex due to the reduction of
the induced paring amplitude in the dot arising from the repulsive Coulomb
interaction. As a consequence the conductance decreases for increasing
$U$ even in this case. This decrease is illustrated in Fig. 3a
where we plot the conductance as a function of $U$ in the symmetric case
for different values of $\Gamma/\Delta$. For large $U/\Gamma$ we find
that the conductance decreases roughly as $(\Gamma/U)^4$. This behavior
can be understood as follows: in order to have a vanishing pairing
amplitude in the $U/\Gamma \rightarrow \infty$ limit, the non-diagonal
self-energy $\Sigma_{12}$ should tend to cancel the 
non-diagonal tunneling rate $(\hat{\Gamma}_S)_{12}$. By analyzing the
expression of diagram d in Fig. 1, this requires that $G_{12}$ decays
as $(\Gamma/U)^2$ and therefore the conductance given by Eq. 
(\ref{conductance}) in our approximation should decay roughly as 
$(\Gamma/U)^4$. This decay is probably less pronounced than in the
exact solution where one would rather expect an exponential  
behavior in the Kondo regime. 

Although the previous analysis shows that the maximum value for the
conductance $4 e^2/h$ can never be reached in the symmetric case for
finite $U$, this is not necessarily the case for an asymmetric
situation with $\Gamma_L \ne \Gamma_R$. In fact, if the coupling to the 
electrodes could be tunned in order to reach the condition
$\tilde{\Gamma}_R = \Gamma_L$ 
then, Eq. (\ref{conductance}) predicts a maximum in the value of $G$.
As shown in Fig. 3b, this condition can be reached by increasing the
coupling to the superconducting electrode. The ratio between $\Gamma_R$
and $\Gamma_L$ at the maximum becomes larger for increasing $U$. 
In a situation with electron hole-symmetry, like the one depicted in
Fig. 3b, the conductance at zero temperature
reaches its maximum possible value $4e^2/h$.

In normal quantum dots a signature of the Kondo effect is given by an
anomalous temperature dependence in the linear conductance 
\cite{Goldhaber}, which exhibits a continuous transition from a maximum
conductance in the Kondo regime to well resolved conductance peaks 
associated with Coulomb blockade. When one of the electrodes is
superconducting there is also a decrease of conductance with temperature
in the Kondo regime. However, as depicted in Fig. 4, the conductance
already exhibits a double peaked structure at zero temperature when
$\Gamma_L = \Gamma_R$. The reduction of conductance with temperature is
in this case much faster than in the normal case, as shown in Fig. 4
(inset). This difference is a consequence of the lowering of the Kondo
temperature due to the presence of the superconducting electrode.

In conclusion, we have analyzed the electronic transport properties
of a quantum dot coupled to a normal and a superconducting lead. For
this purpose we have introduced an electron self-energy which
interpolates between the limits of weak and strong coupling to the leads,
an approach which has been previously used for normal systems
\cite{us,Alvaro,Saso,us2,Alvaro2,Kotliar}. This approximation allows to
describe a broad range of parameters including the relevant one for an
actual experiment. 
On the other hand, we have shown
that for finite charging energy the dot conductance can either be
enhanced or suppressed with respect to the normal case. While in a
symmetrically coupled dot ($\Gamma_L = \Gamma_R$)
an increasing charging energy tends to reduce the
conductance, in the asymmetric case it is always
possible to reach a maximum in the conductance 
by fine tunning the coupling to the superconducting electrode. 
In the case of electron-hole symmetry this maximum reaches the value
$4e^2/h$ at zero temperature. The
predictions presented in this work could be tested experimentally using
similar technologies to those currently used for normal quantum dots
\cite{Goldhaber,Kowenhoven,Takayanagi}  

\acknowledgements
We thank Jan von Delft, Hans Kroha, Andrei Zaikin and Gerd Sch\"on for 
fruitful discussions. This work has been supported by the Spanish CICYT 
under contract No. PB97-0044 and by the SFB 195 of the German Science Foundation.

\begin{figure}[!th]
\begin{center}
\leavevmode
\epsfysize=15cm
\epsfbox{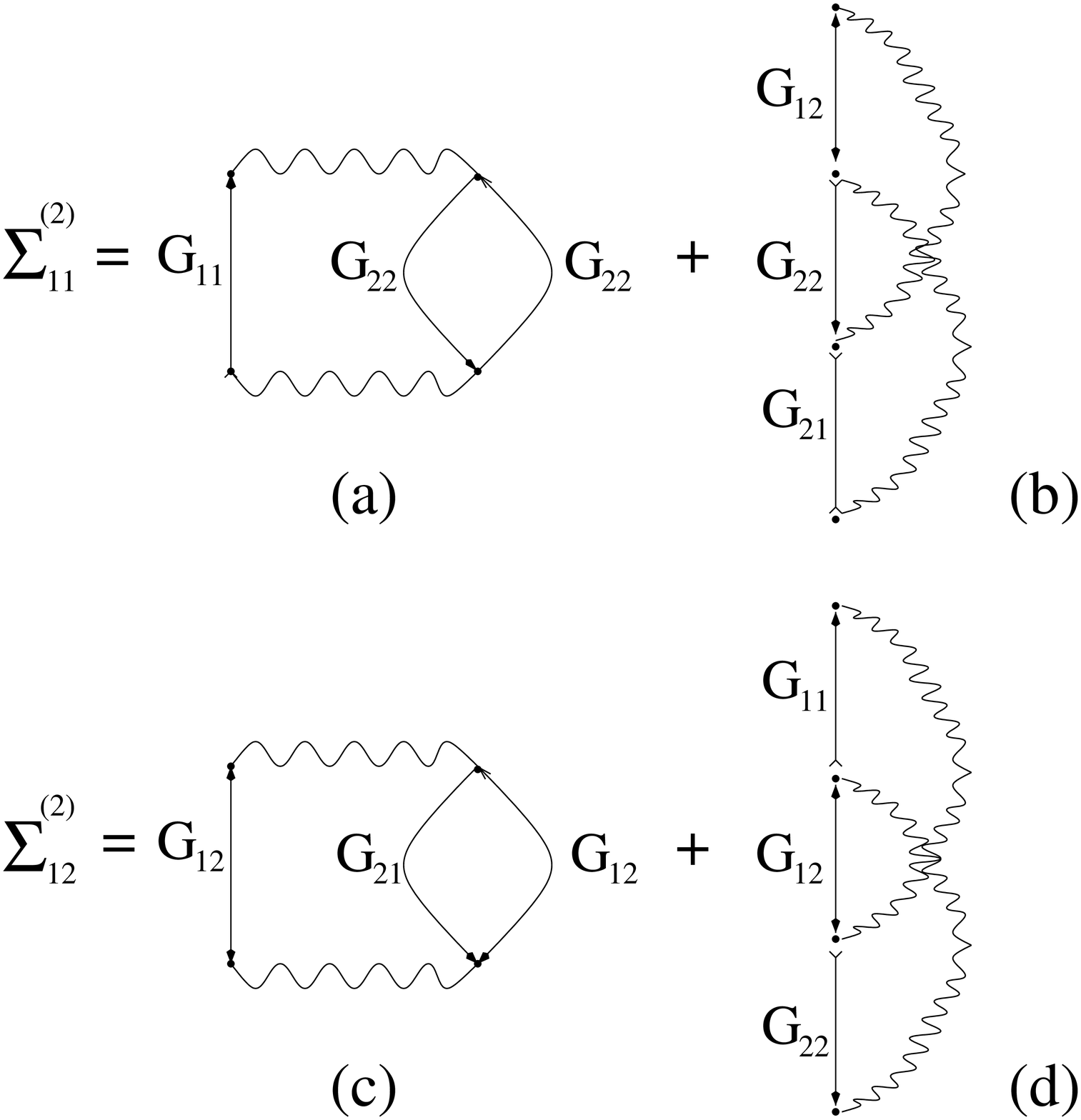}
\end{center}
\caption[]{Second order self-energy diagrams.}
\label{fig1}
\end{figure}

\begin{figure}[!th]
\begin{center}
\leavevmode
\epsfysize=18cm
\epsfbox{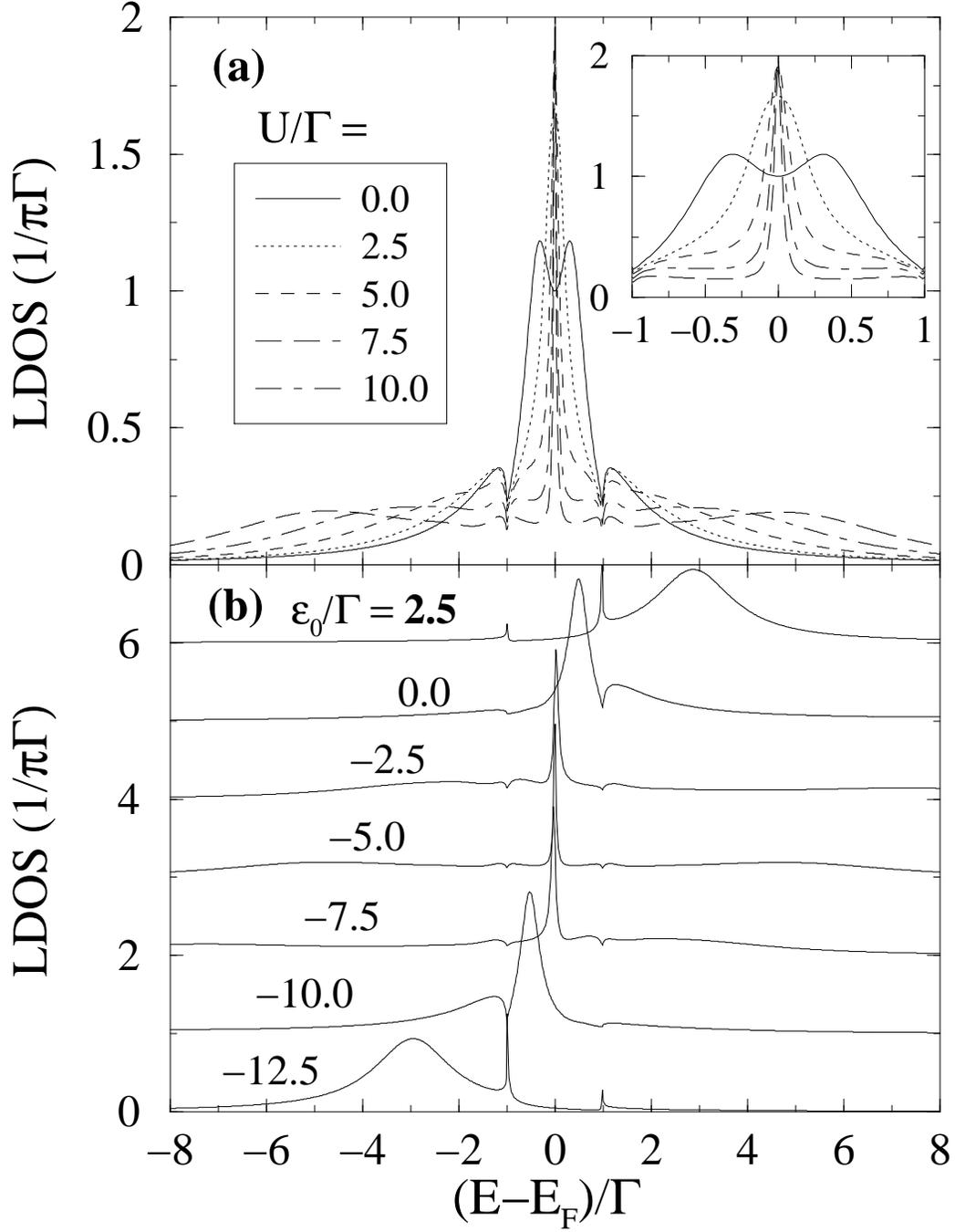}
\end{center}
\caption[]{a) Dot spectral density in the symmetric case ($\epsilon_0 =
-U/2$ and $\Gamma_L = \Gamma_R$) with $\Gamma=\Delta$ for different values of 
$U/\Gamma$. The inset shows a blow up of the region around the Fermi energy. 
b) Dot spectral density for 
different values of $\epsilon_0$ with $U=10 \Gamma$ and $\Gamma = \Delta$.}
\label{fig2}
\end{figure}

\vspace*{3cm}
\begin{figure}[!th]
\begin{center}
\leavevmode
\epsfysize=8cm
\epsfbox{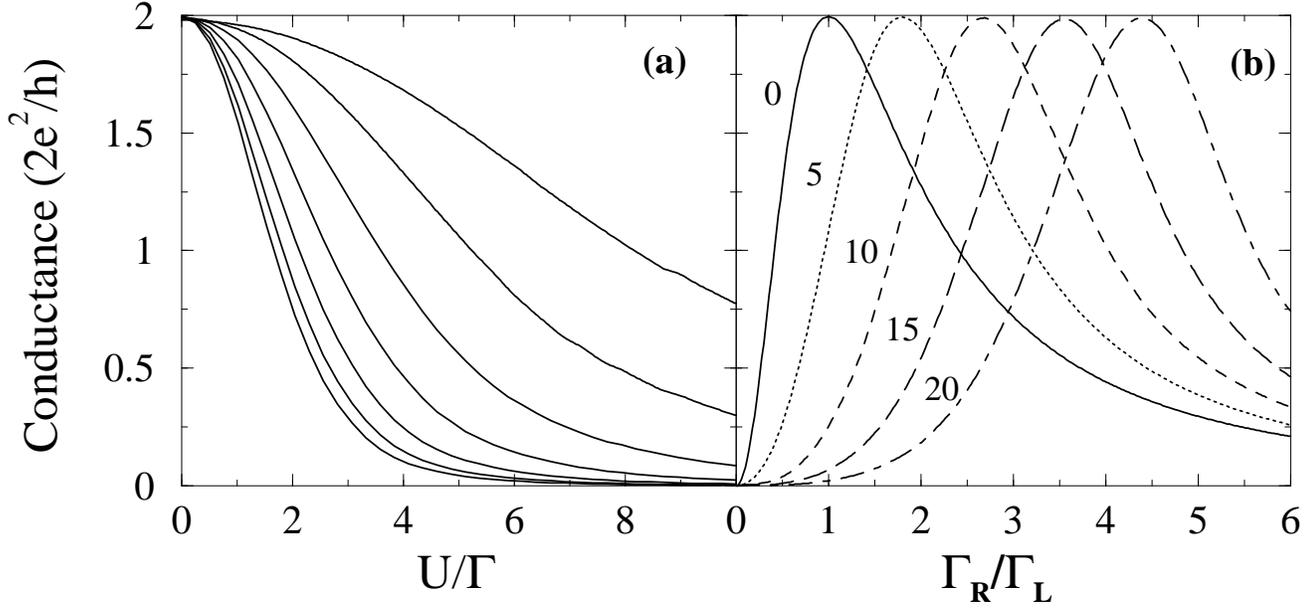}
\end{center}
\caption[]{a) Conductance at zero temperature for the symmetric case as a
function of $U/\Gamma$ and for different values of $\Gamma/\Delta$.
From bottom to top $\Gamma/\Delta =$ 0.125, 0.25, 0.5, 1.0, 2.0, 4.0, 8.0.
b) same as (a) for asymmetric coupling to the leads as a function of
$\Gamma_R/\Gamma_L$ and different values of $U/\Gamma_L$.} 
\label{fig3}
\end{figure}

\vspace*{2cm}
\begin{figure}[!th]
\begin{center}
\leavevmode
\epsfysize=13cm
\epsfbox{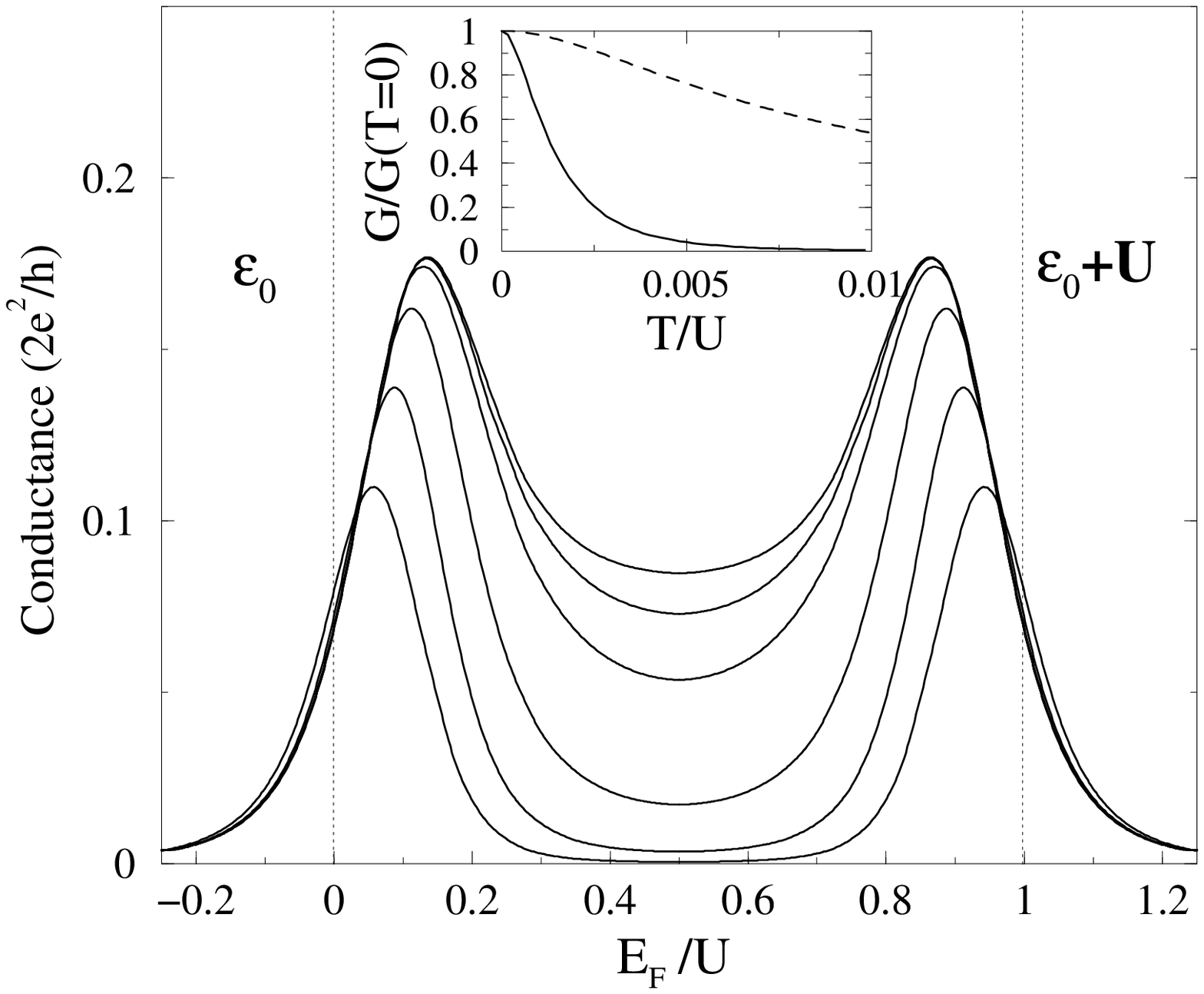}
\end{center}
\caption[]{Conductance for different temperature values: $U=10 \Gamma$,
$\Delta = \Gamma/2$ and $T/U =$ 0.0, 0.0005, 0.001, 0.0025, 0.005, 0.01. Inset:
normalized conductance as a function of temperature for N-dot-S (full line) 
and N-dot-N (dashed line) at $E_F = U/2$.} 
\label{fig4}
\end{figure}


\begin{references}
\bibitem{comment} Moreover, this could be the case of an actual
experimental set up obtained by combining the semiconducting quantum dots
used in Ref. \cite{Goldhaber}, where $\Gamma$ was estimated to be of the
order of $0.1 meV$, with superconducting leads. Notice that the
energy gap in a traditional superconductor like $Al$ would be of the
same order.
\end{references}
\end{document}